\documentclass{jpsj2}

\title{%
Anomalous Enhancement of the Boltzmann Conductivity\\
in Disordered Zigzag Graphene Nanoribbons
}

\author{%
Yositake {\sc Takane}
}

\inst{%
Department of Quantum Matter, Graduate School of Advanced Sciences of Matter,\\
Hiroshima University, Higashihiroshima, Hiroshima 739-8530, Japan
}

\recdate{ \hspace{50mm} }

\abst{%
We study the conductivity of disordered zigzag graphene nanoribbons
in the incoherent regime by using the Boltzmann equation approach.
The band structure of zigzag nanoribbons contains two energy valleys,
and each valley has an excess one-way channel.
The crucial point is that the numbers of conducting channels
for two propagating directions are imbalanced in each valley
due to the presence of an excess one-way channel.
It was pointed out that as a consequence of this imbalance,
a perfectly conducting channel is stabilized in the coherent regime
if intervalley scattering is absent.
We show that even in the incoherent regime, the conductivity is anomalously
enhanced if intervalley scattering is very weak.
Particularly, in the limit of no intervalley scattering, the dimensionless
conductance approaches to unity with increasing ribbon length
as if there exists a perfectly conducting channel.
We also show that anomalous valley polarization of electron density
appears in the presence of an electric field.
}

\kword{%
graphene nanoribbon, perfectly conducting channel, incoherent transport,\\
Boltzmann equation
}

\begin{document}
\sloppy
\maketitle

\section{Introduction}

Graphene nanoribbons with zigzag edges (zigzag nanoribbons) have
a unique band structure~\cite{fujita} which is not seen
in ordinary quantum wires realized in semiconductor nanostructures.
Its band structure contains two valleys, called $K_{-}$ valley and
$K_{+}$ valley, well separated in momentum space, and each valley has
an excess one-way channel arising from a partially flat band.
The crucial point is that the numbers of conducting channels
for two propagating directions are imbalanced in each valley
due to the presence of an excess one-way channel.
This causes an unusual transport property of disordered zigzag nanoribbons
in the coherent transport regime at low temperatures.
Wakabayashi \textit{et al.}~\cite{wakabayashi1,wakabayashi2} have pointed
out that if impurity potentials are long-ranged
and thus intervalley scattering is absent,
one perfectly conducting channel without backward scattering is stabilized
due to the imbalance between the numbers of conducting channels.
This results in the absence of Anderson localization.
Inspired by this observation, the statistical behavior of the conductance
in disordered wires with channel-number imbalance
has been studied extensively.~\cite{takane1,takane2,hirose,takane3,takane4,
takane5,kobayashi,takane6}
The presence of a perfectly conducting channel in such systems
has been suggested by Barnes \textit{et al.}~\cite{barnes1,barnes2}
more than a decade ago.
However, since realistic systems has been lacking,
detailed studies on this subject were not performed until recently.

In addition to zigzag nanoribbons, a perfectly conducting channel appears
in disordered carbon nanotubes.~\cite{ando1,nakanishi,ando2}
In the case of carbon nanotubes, it is stabilized due to the two facts
that the reflection matrix has the skew-symmetry and
the number of conducting channels is odd.~\cite{ando2,takane7,sakai}
Because the skew-symmetry plays a role only in the presence of
the phase coherence of electrons,
we expect that a perfectly conducting channel is fragile against dephasing.
Indeed, Ando and Suzuura~\cite{ando2} showed that a perfectly conducting
channel and related anomalies disappear in the incoherent regime
where the phase coherence of electrons is completely lost by strong dephasing.
Recently, a quasi-perfectly conducting channel has been shown to appear
in disordered graphene nanoribbons with armchair edges.~\cite{yamamoto}

Note that the presence of a perfectly conducting channel
in zigzag nanoribbons is concluded from the fact
that the reflection matrix has a nonsquare form with dimensions
$(N+1) \times N$ due to the imbalance between the numbers of conducting
channels,~\cite{wakabayashi1,hirose,barnes1}
where $N+1$ ($N$) is the number of incident (reflection) channels.
From this fact we can show that one reflection eigenvalue becomes
zero and thus one transmission eigenvalue becomes unity.
This results in the presence of a perfectly conducting channel.
Because the reflection matrix has no special symmetry in this case,
we expect that a perfectly conducting channel is robust against weak dephasing.
However, it should be noted that the reflection matrix itself is no longer
well-defined in the incoherent regime because a single-particle description
of quantum mechanical electron scattering
is not justified under strong dephasing.
That is, the above argument cannot be applied to the incoherent regime,
and therefore we cannot conclude the presence of a well-defined perfectly
conducting channel.
It is of interest and of significance to study
whether anomalous electron transport properties arise
in the incoherent regime.

In this paper we show that even in the incoherent regime, the conductivity
of disordered zigzag nanoribbons shows anomalous behaviors
if impurity potentials are long-ranged
and thus intervalley scattering is very weak.
We introduce two parameters $\kappa$ and $\kappa'$ representing
the strengths of intravalley scattering and intervalley scattering,
respectively, and analytically obtain the conductivity as a function of
$\kappa'/\kappa$ by using the Boltzmann equation approach.
Note that $\kappa' \ll \kappa$ if impurity potentials are long-ranged
and $\kappa' \approx \kappa$ if impurity potentials are short-ranged.
We show that the conductivity is strongly enhanced if $\kappa'/\kappa \ll 1$,
while such an enhancement disappears in the case of $\kappa'/\kappa =1$.
Particularly, in the no intervalley scattering limit of
$\kappa'/\kappa \to 0$, the dimensionless conductance approaches to unity
with increasing ribbon length
as if there exists a perfectly conducting channel.
We also show that anomalous valley polarization of electron density
appears in the presence of an electric field when $\kappa' \ll \kappa$.
We set $\hbar = 1$ in the following.

\section{Analysis Based on the Boltzmann Equation}

\begin{table}[bth]
\caption{The number $N_{\rm R}$ of right-moving channels
and the number $N_{\rm L}$ of left-moving channels in each valley.
}
\begin{center}
\begin{tabular}{c|c|c}
\hline
Valley & $N_{\rm R}$ & $N_{\rm L}$ \\
\hline
$K_{-}$ & $N+m$ & $N$ \\
$K_{+}$ & $N$ & $N+m$ \\
\hline
\end{tabular}
\end{center}
\label{t1}
\end{table}
\begin{figure}[btp]
\begin{center}
\includegraphics[height=6cm]{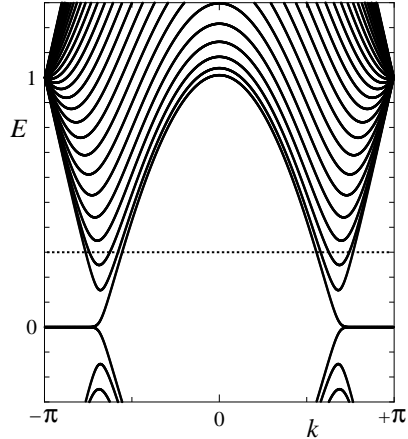}
\end{center}
\caption{The band structure of a zigzag nanoribbon consisting
of 30 zigzag chains.
The left and right valleys correspond to $K_{-}$ and $K_{+}$ valleys,
respectively.
}
\end{figure}
We consider zigzag nanoribbons in which the $K_{-}$ valley has
$N+m$ right-moving channels and $N$ left-moving channels,
and the $K_{+}$ valley has $N$ right-moving channels and
$N+m$ left-moving channels (see Table~I).
Note that $m = 1$ in zigzag nanoribbons,
while $N$ can be controlled by carrier doping.
The case of $m \ge 2$ may be realized in multilayer
zigzag nanoribbons.~\cite{takane1}
Figure~1 shows the band structure of a zigzag nanoribbon
consisting of 30 zigzag chains.
If the Fermi level is at the point indicated by the horizontal dotted line,
the numbers of conducting channels
are characterized by $N = 2$ and $m = 1$.
Hereafter, we abbreviate the right-moving channels
in the $K_{\pm}$ valley as ${\rm R}^{\pm}$
and the left-moving channels in the $K_{\pm}$ valley as ${\rm L}^{\pm}$,
and refer to ${\rm R}^{-}$ and ${\rm L}^{+}$ as majority channels
and ${\rm L}^{-}$ and ${\rm R}^{+}$ as minority channels.
Let $\varepsilon_{{\rm R}nk}^{\pm}$ ($\varepsilon_{{\rm L}nk}^{\pm}$) be
the energy of an electron state with wave number $k$ in the $n$th right-moving
(left-moving) channel of the $K_{\pm}$ valley.
We introduce the corresponding distribution function $g_{Xnk}^{z}$,
where $z = +,-$ and $X = {\rm R}, {\rm L}$.
The group velocity $v_{Xnk}^{z}$ is given by
$v_{Xnk}^{z} = \partial\varepsilon_{Xnk}^{z}/\partial k$.
For simplicity we assume that $v_{{\rm R}nk}^{z} = v_{\rm R}$
and $v_{{\rm L}nk}^{z} = v_{\rm L}$
with $v \equiv v_{\rm R} = -v_{\rm L} > 0$.

We consider a long zigzag nanoribbon of length $L$ placed along the $x$ axis
and apply a constant electric field $E$ in the negative $x$ direction.
This field accelerates electrons in ${\rm R}^{\pm}$
while electrons in ${\rm L}^{\pm}$ are decelerated.
We express the distribution function as
\begin{align}
    \label{eq:g_assumption}
  g_{Xnk}^{z}
   = f_{\rm FD}(\varepsilon_{Xnk}^{z})
     - {\rm sign}(v_{X}) eE \, l_{Xnk}^{z}
       \frac{\partial f_{\rm FD}}{\partial \varepsilon}
       (\varepsilon_{Xnk}^{z}) ,
\end{align}
where $f_{\rm FD}(\varepsilon)$ is the Fermi-Dirac function
and $l_{Xnk}^{z}$, characterizing a deviation from the equilibrium
distribution, is called the mean free path.
This expression implicitly assumes that the applied electric field increases
(decreases) the population in ${\rm R}^{\pm}$ (${\rm L}^{\pm}$).
The distribution function obeys the Boltzmann equation,~\cite{akera}
\begin{align}
      \label{eq:Boltzmann_eq}
   \frac{\partial g_{Xnk}^{z}}{\partial t}
   + v_{X}\frac{\partial g_{Xnk}^{z}}{\partial x}
   + eE \frac{\partial g_{Xnk}^{z}}{\partial k}
   = \sum_{X'n'k'z'}
     W_{Xnk,X'n'k'}^{z,z'}\left(g_{X'n'k'}^{z'}-g_{Xnk}^{z} \right) ,
\end{align}
where $W_{Xnk,X'n'k'}^{z,z'}$ is the scattering probability
between the state with $\{X'n'k'z'\}$ and that with $\{Xnkz\}$.
The scattering probability is expressed as
\begin{align}
  W_{Xnk,X'n'k'}^{z,z'}
    = 2\pi M_{Xnk,X'n'k'}^{z,z'}
      \delta\left(\varepsilon_{X'n'k'}^{z'}-\varepsilon_{Xnk}^{z}\right)
\end{align}
with
\begin{align}
  M_{Xnk,X'n'k'}^{z,z'}
  = \Big\langle \left|U_{Xnk,X'n'k'}^{z,z'}\right|^{2}\Big\rangle ,
\end{align}
where $U_{Xnk,X'n'k'}^{z,z'}$ is the matrix element of impurity potential
and $\langle \cdots \rangle$ indicates the ensemble average over
impurity configurations.
For simplicity we assume that the scattering probability is determined by
only the valley indexes $z$ and $z'$,
and does not depend on details of initial and final states.
Hence $M_{Xnk,X'n'k'}^{\pm,\pm} = M$ and
$M_{Xnk,X'n'k'}^{\pm,\mp} = M'$, where $M$ and $M'$ describe
intravalley scattering and intervalley scattering, respectively.
The Boltzmann conductivity $\sigma$ is obtained as
\begin{align}
    \label{eq:boltzmann-cond}
 \sigma & = \frac{e^{2}}{L}\sum_{Xnkz} |v_{X}|l_{Xnk}^{z}
            \left( - \frac{\partial f_{\rm FD}}{\partial \varepsilon}
                     (\varepsilon_{Xnk}^{z}) \right)
              \nonumber \\
        & = \frac{e^{2}}{2\pi}
            \left( \sum_{n=1}^{N+m}l_{Rn}^{-}+\sum_{n=1}^{N}l_{Ln}^{-}
                 + \sum_{n=1}^{N}l_{Rn}^{+}+\sum_{n=1}^{N+m}l_{Ln}^{+}
            \right) ,
\end{align}
where $l_{Xn}^{z}$ is the mean free path at the Fermi level.
Here and hereafter the spin degeneracy is ignored.

To uncover anomalous features of zigzag nanoribbons in an electric field,
we focus on steady states with spatial uniformity.
According to the assumptions stated above,
the mean free path becomes independent of $n$.
We thus set $l_{Xn}^{z} = l_{X}^{z}$, where $k$ is also dropped
since we are interested in electron states near the Fermi level.
Furthermore we can set $l \equiv l_{\rm R}^{-} = l_{\rm L}^{+}$
and $l' \equiv l_{\rm L}^{-} = l_{\rm R}^{+}$
from the symmetry of the band structure, where $l$ and $l'$ correspond
to the majority and minority channels, respectively.
This ensures charge neutrality of our system.
Substituting eq.~(\ref{eq:g_assumption}) into eq.~(\ref{eq:Boltzmann_eq})
and ignoring the $t$- and $x$-dependent terms,
we obtain two equations for $l$ and $l'$.
In terms of the parameters $\kappa$ and $\kappa'$
defined by
\begin{align}
   & \kappa = \frac{LM}{v^{2}} ,
         \\
   & \kappa' = \frac{LM'}{v^{2}} ,
\end{align}
the resulting equations are expressed as
\begin{align}
  1 & =  \left[N\kappa+(3N+2m)\kappa'\right] l
       + N(\kappa-\kappa') l' ,
           \\
  1 & =  \left[(N+m)\kappa+(3N+m)\kappa'\right] l'
       + (N+m)(\kappa-\kappa') l .
\end{align}
Note that $\kappa$ and $\kappa'$ represent the strength of
intravalley scattering and that of intervalley scattering, respectively.
We easily obtain
\begin{align}
     \label{eq:l_uniform}
  l & = \frac{1}{2(2N+m)^{2}}
        \left(\frac{4N}{\kappa+\kappa'}+\frac{m}{\kappa'} \right) ,
        \\
     \label{eq:l'_uniform}
 l' & = \frac{1}{2(2N+m)^{2}}
        \left(\frac{4(N+m)}{\kappa+\kappa'}-\frac{m}{\kappa'} \right) .
\end{align}
Substituting eqs.~(\ref{eq:l_uniform}) and (\ref{eq:l'_uniform}) into
eq.~(\ref{eq:boltzmann-cond}), we obtain the Boltzmann conductivity as
\begin{align}
  \sigma
   = \frac{e^{2}}{2\pi}\frac{1}{(2N+m)^{2}}
     \left(\frac{8N(N+m)}{\kappa+\kappa'}+\frac{m^{2}}{\kappa'}\right) .
\end{align}
These results show that anomalous features arise in the case of $m \neq 0$.
Firstly, we observe from eq.~(\ref{eq:l'_uniform}) that
$l_{\rm L}^{-}$ and $l_{\rm R}^{+}$
become negative when $\kappa'$ is sufficiently small.
That is, $g_{{\rm L}nk}^{-}$ increases and $g_{{\rm R}nk}^{+}$ decreases.
This indicates that the distribution functions for the minority channels shift
to the direction opposite to that assumed in eq.~(\ref{eq:g_assumption}).
It should be emphasized that because $g_{Xnk}^{-}$ increases and
$g_{Xnk}^{+}$ decreases regardless of the propagating directions,
electron density is polarized between the two valleys in the presence of
an electric field.
Secondly, we observe from eqs.~(\ref{eq:l_uniform}) and (\ref{eq:l'_uniform})
that all the mean free paths diverge in the limit of $\kappa' \to 0$
as in ballistic quantum wires.
Accordingly, the Boltzmann conductivity also diverges.

To gain an insight into these anomalous features,
we consider the Boltzmann equation
including the $t$-dependent term in the limit of $\kappa' \to 0$.
The two valleys are completely decoupled in this limit,
so we treat only the $K_{-}$ valley.
The Boltzmann equation yields
\begin{align}
     \label{eq:exp_acc1}
  \frac{1}{v}\frac{\partial l_{\rm R}^{-}}{\partial t}
    & = 1 - N\kappa \left(l_{\rm R}^{-}+l_{\rm L}^{-}\right) ,
         \\
     \label{eq:exp_acc2}
  \frac{1}{v}\frac{\partial l_{\rm L}^{-}}{\partial t}
    & = 1 - (N+m)\kappa \left(l_{\rm R}^{-}+l_{\rm L}^{-}\right) .
\end{align}
In the right-hand side of each equation,
the first term represents the acceleration induced by an electric field,
while the second term describes the deceleration due to backward scattering.
These equations indicate that the electric field increases $l_{\rm R}^{-}$
and $l_{\rm L}^{-}$ in the same rate but the deceleration effect for
$l_{\rm R}^{-}$ is by factor $N/(N+m)$ smaller than that for $l_{\rm L}^{-}$
due to the imbalance between the numbers of conducting channels.
Therefore, the acceleration and the deceleration are never balanced
in the case of $m \neq 0$,
and $l_{\rm R}^{-}$ and $l_{\rm L}^{-}$ do not converge.
The acceleration dominates the deceleration in ${\rm R}^{-}$
(i.e., majority channels) while their relative strengths are interchanged
in ${\rm L}^{-}$ (i.e., minority channels).
This indicates that  $l_{\rm R}^{-} \to +\infty$ and
$l_{\rm L}^{-} \to -\infty$.
Similarly, we observe that $l_{\rm L}^{+} \to +\infty$ and
$l_{\rm R}^{+} \to -\infty$.
We understand that the anomalous increase of $l$ (i.e., $l \to +\infty$)
is caused by excess electric field effect while the anomalous decrease
of $l'$  (i.e., $l' \to -\infty$) is caused by excess backward scattering.
From the argument given above, the characteristic behavior of electrons
in the limit of $\kappa' \to 0$ with $m \neq 0$ is summarized as follows.
Firstly, $l \to +\infty$ implies that the behavior of electrons in the
majority channels is similar to that in ballistic quantum wires
in spite of the presence of backward scattering due to impurities.
Secondly, $l' \to -\infty$ implies that the behavior of electrons in the
minority channels is also similar to that in ballistic quantum wires
but the electric field effect arises in the direction opposite to
that naturally expected.
That is, electrons behave as if they are free from impurity scattering
but an electric field affects electrons in the minority channels
as if its sign is reversed.

The divergence of the mean free paths must be removed
if we correctly take account of the fact that electrons are accelerated
by an electric field only in a finite region.
Therefore we reconsider the Boltzmann equation
including the $x$-dependent term.
Suppose that a constant electric field $E$ is applied
only in the region of $-L/2 < x < L/2$.
We consider that incident electrons from left (right) are described by
the equilibrium distribution at $x = -L/2$ ($x = L/2$).
This indicates that~\cite{de_Jong}
\begin{align}
      \label{eq:BC1}
   & l_{\rm R}^{-}(-L/2)=l_{\rm R}^{+}(-L/2)=0 ,
        \\
      \label{eq:BC2}
   & l_{\rm L}^{-}(L/2)=l_{\rm L}^{+}(L/2)=0 .
\end{align}
From the symmetry of the band structure, we assume that
\begin{align}
  l(x) & \equiv l_{\rm R}^{-}(x)=l_{\rm L}^{+}(-x) ,
     \\
 l'(x) & \equiv l_{\rm R}^{+}(x)=l_{\rm L}^{-}(-x) .
\end{align}
The Boltzmann equation for steady states yields
\begin{align}
  -\frac{\partial l(x)}{\partial x}+1
    & =  \left[N\kappa+(2N+m)\kappa'\right]l(x) + N\kappa l'(-x)
           \nonumber \\
    &   + (N+m)\kappa'l(-x) - N\kappa'l'(x) ,
         \\
  -\frac{\partial l'(x)}{\partial x}+1
    & =  \left[(N+m)\kappa+(2N+m)\kappa'\right]l'(x) + (N+m)\kappa l(-x)
           \nonumber \\
    &   + N\kappa'l'(-x) - (N+m)\kappa'l(x) .
\end{align}
We solve these coupled equations under the boundary condition,
eqs.~(\ref{eq:BC1}) and (\ref{eq:BC2}), and obtain
\begin{align}
       \label{eq:result_l}
  l(x)
 & = x + \frac{L}{2}
    + \frac{L}{\Sigma}
      \bigg[ - m(\kappa-\kappa')\left( c(x)-c_{0} \right)
             + \frac{m^{2}(\kappa^{2}-{\kappa'}^{2})}{\sqrt{\alpha}}
               \left( d(x)-d_{0} \right)
        \nonumber \\
 & \hspace{30mm}
             - 2(2N+m)^{2}(\kappa+\kappa')\kappa'c_{0}(x+\frac{L}{2})
      \bigg] ,
                  \\
       \label{eq:result_l'}
  l'(x)
 & = x + \frac{L}{2}
    + \frac{L}{\Sigma}
      \bigg[   m(\kappa-\kappa')\left( c(x)-c_{0} \right)
             + \frac{m^{2}(\kappa^{2}-{\kappa'}^{2})}{\sqrt{\alpha}}
               \left( d(x)-d_{0} \right)
        \nonumber \\
 & \hspace{30mm}
             - 2(2N+m)^{2}(\kappa+\kappa')\kappa'c_{0}(x+\frac{L}{2})
      \bigg] ,
\end{align}
where $\alpha=(\kappa+\kappa')[m^{2}\kappa+(8N^{2}+8Nm+m^{2})\kappa']$ and
\begin{align}
  c(x) & = \frac{(2N+m)(\kappa+\kappa')}{\sqrt{\alpha}}
           \cosh(\sqrt{\alpha}x) - \sinh(\sqrt{\alpha}x) ,
             \\
  d(x) & = - \frac{(2N+m)(\kappa+\kappa')}{\sqrt{\alpha}}
           \sinh(\sqrt{\alpha}x) + \cosh(\sqrt{\alpha}x) .
\end{align}
The constants in the above equations are given as follows:
$c_{0}=c(-L/2)$, $d_{0}=d(-L/2)$, and
\begin{align}
  \Sigma = 4(2N+m)\kappa'
           \left(1 + \frac{1}{2}(2N+m)(\kappa+\kappa')L\right)c_{0}
           + 2m^{2}\frac{\kappa^{2}-{\kappa'}^{2}}{\sqrt{\alpha}}d_{0} .
\end{align}

We now obtain the Boltzmann conductivity $\sigma(x)$ and
the two-terminal conductance $G$.
Equation~(\ref{eq:boltzmann-cond}) yields
\begin{align}
     \label{eq:sigma_mod}
  \sigma(x)
  = \frac{e^{2}}{2\pi}
    \left[ (N+m)\left(l(x)+l(-x)\right) + N\left(l'(x)+l'(-x)\right) \right] .
\end{align}
Substituting eqs.~(\ref{eq:result_l}) and (\ref{eq:result_l'})
into eq.~(\ref{eq:sigma_mod}), we find that $\sigma(x)$ is independent of $x$
and is given by
\begin{align}
  \sigma
   = \frac{e^{2}L}{2\pi}
     \frac{2\alpha c_{0}}{(\kappa + \kappa')\Sigma} .
\end{align}
This ensures current continuity in our system.
The conductance $G = \sigma / L$ is expressed as
\begin{align}
      \label{eq:conductance}
  G = \frac{e^{2}}{2\pi}
      \frac{\left[m^{2}\kappa + (8N^{2}+8Nm+m^{2})\kappa' \right]c_{0}}
           {2(2N+m)\kappa'\left[1+\frac{1}{2}(2N+m)(\kappa+\kappa')L
                          \right]c_{0}
            + m^{2}\frac{\kappa^{2}-{\kappa'}^{2}}{\sqrt{\alpha}}d_{0}} .
\end{align}

Let us consider the no intervalley scattering limit of $\kappa' \to 0$,
at which we obtain
\begin{align}
      \label{eq:quantization}
  G = \frac{e^{2}}{2\pi}
      \left(m + \frac{2Nm}{(N+m){\rm e}^{m\kappa L}-N}\right).
\end{align}
We observe that the dimensionless conductance $g \equiv (2\pi/e^{2})G$
behaves as $g \to m$ in the limit of $\kappa L \to \infty$
as if there exist $m$ perfectly conducting channels.
This indicates that the conductance in the limit of $\kappa' \to 0$
is drastically enhanced and greatly deviates from Ohm's law.
We interpret it on the basis of the argument given below
eqs.~(\ref{eq:exp_acc1}) and (\ref{eq:exp_acc2}).
We show that very anomalous electron distributions are induced
by an electric field in the limit of $\kappa' \to 0$ with $m \neq 0$.
These distributions imply that electrons behave as if
they are free from impurity scattering as in ballistic quantum wires
but the electric field effect on ${\rm L}^{-}$ and ${\rm R}^{+}$
arises as if the sign of an electric field is reversed.
This observation indicates that the contribution to $g$ from ${\rm R}^{-}$
and ${\rm L}^{+}$ is $N+m$ while that from ${\rm L}^{-}$ and ${\rm R}^{+}$
is negative and is given by $-N$.
By summing these two contributions we obtain $g = m$.
This roughly accounts for our result in the limit of $\kappa L \to \infty$.

Before considering the case of $\kappa' > 0$,
we briefly examine the two limiting cases
in which Ohmic behavior of $G$ is expected.
In the absence of one-way excess channels (i.e., $m = 0$),
eq.~(\ref{eq:conductance}) is reduced to
\begin{align}
  G = \frac{e^{2}}{2\pi}
      \frac{2N}{1+N(\kappa+\kappa')L} ,
\end{align}
which obeys Ohm's law when $N(\kappa+\kappa')L \gg 1$.
We next consider the case of $\kappa' = \kappa$
in which the two valleys are completely connected
and the imbalance between the numbers of conducting channels disappears.
In this case, eq.~(\ref{eq:conductance}) is reduced to
\begin{align}
  G = \frac{e^{2}}{2\pi}
      \frac{2N+m}{1+(2N+m)\kappa L} ,
\end{align}
which also obeys Ohm's law when $(2N+m)\kappa L \gg 1$.

We now consider how the behavior of the conductance
is affected by $\kappa'$ in the case of $m \neq 0$.
We numerically obtain the dimensionless conductance $g = (2\pi/e^{2})G$
as a function of $\kappa L$ or $\kappa'/\kappa$.
Figure~2 shows $g$ as a function of $\kappa L$
for $\kappa'/\kappa = 0.0001$, $0.01$, and $1.0$
in the case of $N = 2$ and $m = 1$.
\begin{figure}[btp]
\begin{center}
\includegraphics[height=5cm]{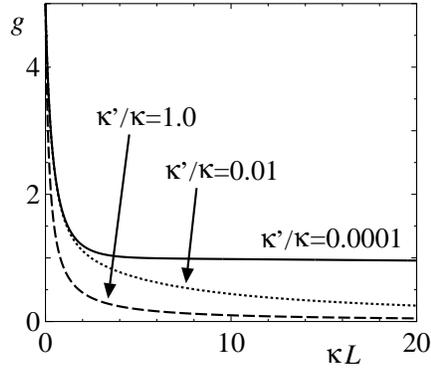}
\end{center}
\caption{The dimensionless conductance as a function of $\kappa L$
for $\kappa'/ \kappa = 0.0001$, $0.01$, and $1.0$
in the case of $N = 2$ and $m = 1$.
}
\end{figure}
We observe that $g$ is nearly unity even when $\kappa L = 20$
in the case of $\kappa'/\kappa = 0.0001$.
Figure~3 shows $g$ as a function of $\kappa'/\kappa$
for $\kappa L = 10$ and $20$ in the case of $N = 2$ and $m = 1$.
Since the behavior of $g$ for small $\kappa'$ is of interest,
we treat only the region of $\kappa'/\kappa \le 0.01$.
The result for the case of $N = 3$ and $m = 0$ is also displayed
for comparison.
\begin{figure}[btp]
\begin{center}
\includegraphics[height=6cm]{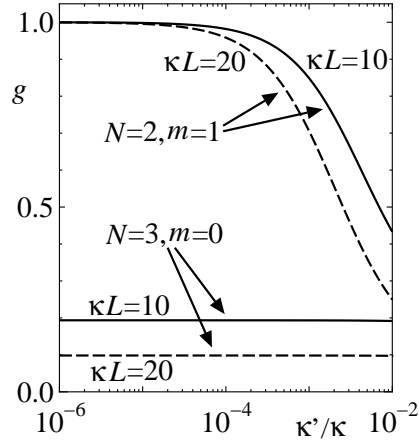}
\end{center}
\caption{The dimensionless conductance as a function of $\kappa '/\kappa$
for $\kappa L= 10$ and $20$ in the case of $N = 2$ and $m = 1$,
and that in the case of $N = 3$ and $m = 0$.
}
\end{figure}
We observe that in the case of $N = 2$ and $m = 1$, $g$ approaches to unity
with decreasing $\kappa'/\kappa$, while $g$ in the case of $N = 3$ and $m = 0$
is nearly independent of $\kappa'/\kappa$.
Furthermore, $g$ in the case of $N = 2$ and $m = 1$ is much larger than
that in the case of $N = 3$ and $m = 0$
in spite of the fact that the total number of conducting channels
in the latter case is by one greater than that in the former case.
This indicates that the conductance is strongly enhanced
in the case of $m \neq 0$ if $\kappa'$ is sufficiently small
and thus the intervalley scattering is very weak.

\section{Summary and Discussion}

In summary, we have studied the conductivity of disordered zigzag graphene
nanoribbons in the incoherent regime by using the Boltzmann equation approach.
We have shown that the Boltzmann conductivity of zigzag nanoribbons
is anomalously enhanced if intervalley scattering is very weak.
Particularly, in the limit of no intervalley scattering, the dimensionless
conductance approaches to unity with increasing ribbon length
as if there exists a perfectly conducting channel.
We have also shown that anomalous valley polarization of electron density
appears in the presence of an electric field
if intervalley scattering is very weak.

We here consider how the dimensionless conductance $g$ of disordered zigzag
graphene nanoribbons behaves with increasing temperature,
focusing on the effect of dephasing.
For simplicity we restrict our consideration to
the most interesting limit of no intervalley scattering.
At zero temperature where the effect of dephasing can be ignored,
we observe that one channel becomes perfectly conducting and the other channels
are subjected to Anderson localization.~\cite{wakabayashi1,wakabayashi2}
Hence $g$ decreases exponentially toward unity with increasing ribbon length.
The effect of dephasing becomes stronger with increasing temperature.
However, as long as dephasing is weak, we expect that one perfectly
conducting channel survives as suggested in \S 1
and therefore $g$ behaves as in the zero temperature limit.
Our result eq.~(\ref{eq:quantization}) indicates that the behavior of $g$
is almost unchanged even if temperature is further increased
and the dephasing effect becomes strong.
This should be contrasted to the behavior of $g$
in disordered carbon nanotubes in which
a perfectly conducting channel and related anomalies are
strongly suppressed by dephasing.~\cite{ando2}
We conclude that the temperature range in which the anomalous enhancement
of the conductivity can be observed is much wider in zigzag nanoribbons
than in carbon nanotubes.

\section*{Acknowledgment}

This work was supported in part by a Grant-in-Aid for Scientific
Research (C) (No. 21540389)
from the Japan Society for the Promotion of Science.

\end{document}